\documentclass[]{aa}

\usepackage{graphicx}
\usepackage{txfonts}

\begin{document}

\title{A population synthesis study of  the luminosity function of hot
  white dwarfs}

\author{S. Torres\inst{1,2},
        E. Garc\'{\i}a--Berro\inst{1,2},
        J. Krzesinski\inst{3} \and 
        S. J. Kleinman\inst{4}}
\institute{Departament de F\'\i sica Aplicada, 
           Universitat Polit\`ecnica de Catalunya,  
           c/Esteve Terrades, 5,  
           08860 Castelldefels,  
           Spain 
           \and
           Institute for Space Studies of Catalonia,
           c/Gran Capit\`a 2--4, Edif. Nexus 104,   
           08034  Barcelona, 
	   Spain 
           \and
           Mt. Suhora Observatory, 
           Cracow Pedagogical University, 
           ul. Podchorazych 2, 
           30-084 Cracow, 
           Poland 
           \and
           Gemini Observatory, 
           670 N. A'Ohoku Place, 
           Hilo, 
           HI 96720, 
           USA}

\date{\today}

\offprints{E. Garc{\'\i}a--Berro}

\abstract{We present a coherent and detailed Monte Carlo simulation of
           the  population   of  hot  white  dwarfs.   We  assess  the
           statistical significance of the hot  end of the white dwarf
           luminosity function  and the role played  by the bolometric
           corrections of hydrogen-rich white dwarfs at high effective
           temperatures.}
         {We use  the most up-to-date stellar  evolutionary models and
           implement a full description of the observational selection
           biases  to obtain  realistic  simulations  of the  observed
           white dwarf population.}
         {Our  theoretical results  are compared  with the  luminosity
           function  of  hot  white  dwarfs obtained  from  the  Sloan
           Digital Sky  Survey (SDSS),  for both  DA and  non-DA white
           dwarfs.}
        {We  find  that  the  theoretical  results  are  in  excellent
          agreement with the observational  data for the population of
          white  dwarfs with  hydrogen  deficient atmospheres  (non-DA
          white  dwarfs).  For the  population  of  white dwarfs  with
          hydrogen-rich atmospheres  (white dwarfs  of the  DA class),
          our   simulations   show   some   discrepancies   with   the
          observations for the brightest luminosity bins, namely those
          corresponding  to   luminosities  larger  than   $\sim  10\,
          L_{\sun}$. These discrepancies can  be attributed to the way
          in which the masses of the white dwarfs contributing to this
          luminosity  bin have  been computed,  as most  of them  have
          masses  smaller   than  the  theoretical  lower   limit  for
          carbon-oxygen white dwarfs.}
         {We  conclude  that  the   way  in  which  the  observational
           luminosity function of hot white dwarfs is obtained is very
           sensitive to  the particular  implementation of  the method
           used to derive the masses of  the sample. We also provide a
           revised  luminosity  function  for hot  white  dwarfs  with
           hydrogen-rich atmospheres.}


\titlerunning{The luminosity function of hot white dwarfs}
\authorrunning{S. Torres et al.}

\maketitle


\section{Introduction}

White dwarfs are  the most common end-point of  stellar evolution.  In
fact, more than  $\sim 95\%$ of the stars in  our Galaxy, namely those
with masses smaller than $\sim 10\, M_{\sun}$, will end their lives as
white dwarfs. Hence, the white  dwarf population carries very valuable
information about the  evolution of the vast majority of  stars in our
Galaxy.  Moreover,  their structural  and evolutionary  properties are
relatively well understood -- see,  for instance, the recent review of
\citet{Althausetal10}.   Consequently,  the characteristics  of  white
dwarf populations can be used  to constrain the evolutionary models of
their progenitor stars.  For instance,  the white dwarf population can
be used  to assess  the amount  of mass lost  by their  progenitors, a
process that  plays a  key role  in stellar  evolution.  Additionally,
since the  mass ejected  in the  late stages  of stellar  evolution is
enriched  in heavy  elements, the  study of  the Galactic  white dwarf
population is important to model the chemical evolution of our Galaxy.
These, however, are  not the only useful applications  of studying the
population of white dwarfs.  It is  worth mentioning that the study of
the local  white dwarf population provides  interesting constraints on
the local star  formation rate \citep{DPetal94,Rowell} and  on the age
of    the     Galactic    disk     in    the     Solar    neighborhood
\citep{winget,nature,GB88} and  other Galactic  populations, including
open  \citep{nature2}  and  globular  clusters --  see,  for  example,
\cite{hansen07}.   Furthermore, the  ensemble  characteristics of  the
population of white dwarfs can be used to constrain the mass of weakly
interacting  particles, like  axions \citep{Isern},  neutrino emission
rates  \citep{lambvh75} and  the nature  of the  baryonic dark  matter
content of our Galaxy -- see, for instance, \citet{halo}, \citet{MC2},
and \citet{MC3},  and references therein.  Finally,  the population of
white dwarfs can also be used to study alternative theories of gravity
\citep{JCAP,gdot}.

All these investigations require  not only accurate evolutionary white
dwarf  cooling  sequences, but  also  reliable  determinations of  the
observed ensemble properties of the white dwarf population, namely its
luminosity function and  its mass distribution.  Thus,  to address all
these  questions,  we   need  large  white  dwarf   samples  of  known
completeness.   The Sloan  Digital  Sky Survey  (SDSS)  -- see,  e.g.,
\cite{yor00} --  has provided  us with  a very  large sample  of white
dwarfs and from it, a reliable white dwarf luminosity function for hot
white dwarfs -- namely those  with effective temperatures in excess of
23,500~K --  has been  recently obtained \citep{Krz2009}.   This white
dwarf luminosity function has the interesting benefit that it has been
obtained  using solely  spectroscopically-confirmed white  dwarfs from
the SDSS DR4 and it is thus an excellent testbed to check not only the
white  dwarf cooling  sequences  at high  luminosities,  but also  our
ability to  model reliably some  Galactic inputs necessary  to compute
the  luminosity  function.    Here  we  describe  the   results  of  a
comprehensive set  of Monte Carlo  simulations aimed to model  the hot
part of  the white  dwarf luminosity  function for  both hydrogen-rich
(DA) white dwarfs and non-DA (hydrogen deficient) stars.

\begin{figure}[t]
\vspace{8.5cm}    
\includegraphics{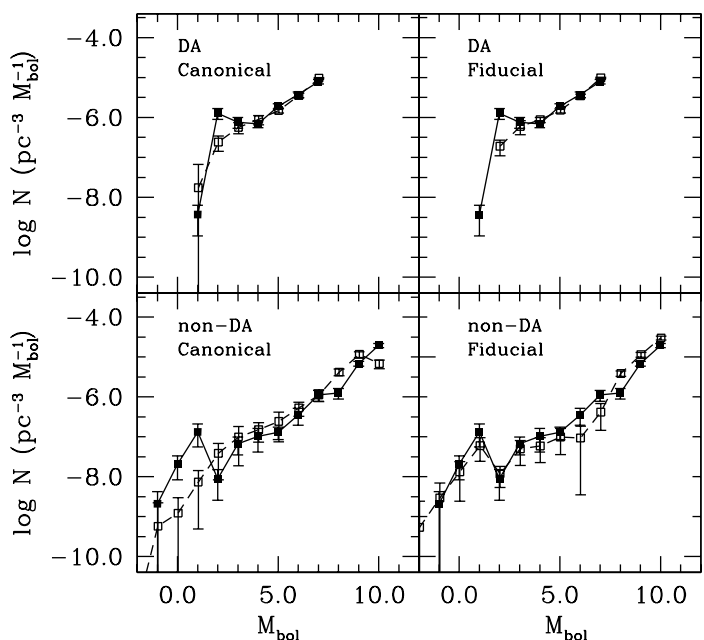}
\caption{White dwarf luminosity functions for  hot DA and non-DA white
  dwarfs for two  different models -- see  text Sect.~\ref{sec:LF} for
  details.   The solid  line  with filled  squares  is the  luminosity
  function  of \citet{Krz2009},  while  the dashed  lines with  hollow
  squares show the results of our simulations.}
\label{f:hotlf}
\end{figure}

\section{The population synthesis code}

Our  population  synthesis  code  has been  extensively  described  in
previous  papers \citep{MC1,MC2,MC3},  and  has proven  to  be a  very
valuable  tool to  simulate successfully  the Galactic  disk and  halo
populations, as well  as the populations of  Galactic clusters. Hence,
we will only describe here its most relevant aspects, and we refer the
interested reader  to previous publications for  detailed descriptions
of the  relevant physical  and astronomical inputs.   Specifically, we
simulated a  synthetic population  of disk white  dwarfs in  the Solar
neighborhood in a sphere of radius 3~kpc.  This radius is sufficiently
large to  avoid possible biases  due to the sampling  procedure, which
for the case under study could  be important given that we are dealing
with the brightest population of white  dwarfs. We also adopted a disk
age  of 10.5~Gyr,  a  constant  star formation  rate,  and a  standard
initial mass  function \citep{Kroupa_2001}. However, let  us note that
the results presented  in this paper do not sensitively  depend on the
precise choice of the initial  mass function.  In particular, when the
initial mass  function of \cite{Scalo}  is adopted we  find negligible
changes in our  results. Velocities were obtained  taking into account
the differential rotation of the  Galaxy, the peculiar velocity of the
Sun, and a dispersion law which  depends on the Galactic scale height.
We used a  double exponential profile with a  scale height $h=250\,$pc
and  a  scale  length  $l=1.3\,$kpc,  and  adopted  the  initial-final
relationship  of  \citet{cataetal2008}   and  \citet{catab2008}.   The
cooling sequences  employed varied  per the mass  of the  white dwarf,
$M_{\rm  WD}$.   If $M_{\rm  WD}\leq  1.1\,M_{\sun}$,  a CO  core  was
adopted, while if  $M_{\rm WD}>1.1\, M_{\sun}$, an ONe  core was used.
In  the case  of CO  white dwarfs  with H-rich  envelopes we  used the
evolutionary calculations of \citet{Renedo_10}, while for white dwarfs
with ONe cores we used  those of \citet{Althaus2007}.  For H-deficient
white dwarfs, we used  the cooling sequences of \citet{Benvenuto1997},
which  correspond   to  pure   He  atmospheres,  and   the  bolometric
corrections of \citet{Bergeron}.  Finally,  we used the same selection
criteria in our  model population as \citet{Krz2009}  employed to cull
their  observational sample.   Specifically,  we  selected only  white
dwarf models  with $g>14$  and a  fully de-reddened  magnitude $g_{\rm
o}<19$  that satisfied  the  color cuts  -$1.5<(u-g)_{\rm  o}< 0$  and
$-1.5<(g-r)_{\rm o}<0$.

\section{The luminosity function of hot white dwarfs}
\label{sec:LF}

In Fig.~\ref{f:hotlf} we  show the model luminosity  functions for hot
hydrogen-rich (DA)  and hydrogen-deficient  (non-DA) white  dwarfs for
two  different assumptions  about  the  ratio of  DA  to non-DA  white
dwarfs.  Each  model luminosity function  shown in this figure  is the
ensemble average of 50 independent  Monte Carlo realizations for which
we also computed the corresponding  standard deviations as an estimate
of  the associated  sampling errors.  In  the {\sl  canonical} set  of
simulations   we   adopt   the  standard   DA/non-DA   ratio   $f_{\rm
DA/non-DA}=0.80$, independent  of effective temperature.  In  the {\sl
fiducial}  set  of  simulations,  we  use a  ratio  dependent  on  the
effective temperature, $T_{\rm eff}$, adopting the ratio obtained from
the  SDSS  \citep{Krz2009}.   The solid  lines  in  Fig.~\ref{f:hotlf}
represent  the white  dwarf  luminosity  function of  \citet{Krz2009},
while  the  dashed  lines  show our  synthetic  luminosity  functions,
obtained from the  model population of hot white  dwarfs computed with
the procedure outlined  in the previous section.  The  upper panels of
Fig.~\ref{f:hotlf} clearly show that the luminosity function of hot DA
white  dwarfs barely  depends on  the choice  of the  DA/non-DA ratio,
whilst that of  non-DA white dwarfs depends sensitively  on it.  Also,
it is quite apparent that the agreement with the observational data is
excellent for  the fiducial model in  which the ratio of  non-DA to DA
white dwarfs depends on effective temperature.

There is  one puzzling characteristic of  the observational luminosity
function of hot DA white  dwarfs clearly seen in Fig.~\ref{f:hotlf} --
namely,  the  existence of  a  plateau  at luminosities  around  $\log
(L/L_{\sun})\approx 1$, or equivalently  $M_{\rm bol}\approx 2.0$ that
is   not   reproduced   in   our   synthetic   luminosity   functions.
Interestingly, the luminosity function of hot non-DA white dwarfs does
not have  a similar plateau  and there is excellent  agreement between
the  synthetic  and observational  data.   Thus,  the origin  of  this
feature in the DA luminosity function cannot be attributed to a global
Galactic input -- like the initial  mass function or to a recent burst
of star formation -- because if  this were the case, the feature would
be present in the luminosity functions of both the DA and non-DA white
dwarfs.  Also, the  existence of this plateau cannot  be a consequence
of the gross  evolutionary properties of the progenitor  stars -- like
the choice  of the initial-final  mass relationship -- since  again in
this case we  would find a similar feature in  the luminosity function
of non-DA white dwarfs. Instead, the  presence of this plateau must be
related to the  intrinsic way in which the luminosity  function of hot
DA white  dwarfs is derived.  Hence,  we are left with  three possible
alternatives to explain its origin, although other explanations cannot
be obviously discarded ``a priori''.   Either the model atmospheres of
hot  DA white  dwarfs are  not complete  for this  range of  effective
temperatures  and there  is  a  piece of  physics  that  has not  been
properly  taken  into  account  in   the  model  atmospheres,  or  the
theoretical  cooling  sequences are  not  entirely  reliable at  these
luminosities -- possibly  influenced by the initial  conditions of the
cooling models,  or by  the model atmospheres  employed to  derive the
colors  and   the  bolometric  magnitude  --   or  alternatively,  the
observational error bars  inherent to the $1/V_{\rm  max}$ method have
been underestimated  -- a  reasonable possibility  given that  at this
range of luminosities  the number of objects is  relatively small, see
\citet{Geijoetal06} --  and the  plateau corresponds to  a statistical
fluctuation.  We study all three possibilities in detail.

\begin{figure}[t]
\vspace{12.0cm}
\includegraphics{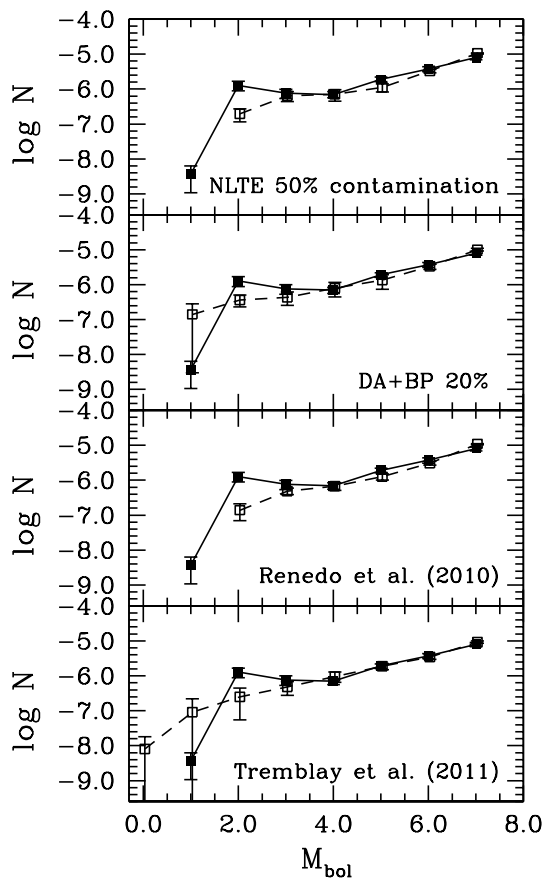}
\caption{Synthetic luminosity  function of  hot white  dwarfs obtained
  using different assumptions about the atmospheric composition of hot
  DA white dwarfs  and two sets of model atmospheres,  compared to the
  observational  luminosity function  of the  SDSS \citep{Krz2009}  --
  solid  line  and  filled  squares.   The top  two  panels  show  the
  synthetic  luminosity functions  of hot  DA white  dwarfs when  NLTE
  corrections  and metal  contamination, and  when the  corrections of
  \cite{Gianinas2010}  to   the  so-called  Balmer-line   problem  are
  considered.   The  two bottom  panels  display  the results  of  our
  simulations  when  the  cooling   tracks  of  \citet{Renedo_10}  are
  employed --  that is, our  fiducial model  -- whereas in  the second
  bottom  panel  we  display  the  results  obtained  when  the  model
  atmospheres of \cite{Tremblay_11} are used.   In all four panels the
  results of  the population synthesis calculations  are plotted using
  dashed lines and open squares.}
\label{f:lf_models}
\end{figure}

In a  first attempt  to explore the  origin of the  plateau in  the DA
white dwarf luminosity function, we  studied the effect of the adopted
atmospheres  in  the  analysis  of  the population  of  hot  DA  white
dwarfs. Specifically, we analyzed the  effects of the NLTE corrections
and of the presence of metals  in the atmospheres of these stars.  For
this    purpose     we    adopted    the    NLTE     corrections    of
\citet{Napi1999}. These corrections are provided in a tabular form for
temperatures  $30,000\,{\rm   K}<T_{\rm  eff}<100,000\,{\rm   K}$  and
gravities $6.50  <\log g<9.75$.  Additionally, we  assumed that 50$\%$
of hot DA white dwarfs have  metals in their atmospheres, although the
results are essentially  the same when different  percentages of white
dwarfs contaminated by  metals are adopted.  According  to the results
of \citet{Napi1999}, the effective temperatures (and consequently, the
luminosities) of  DA white dwarfs with  metal-contaminated atmospheres
are overestimated by  20$\%$ at $80,000\,$K, while  at $40,000\,$K, no
corrections are  needed.  These  effective temperatures  correspond to
approximately the correct range of  luminosities for which the plateau
in the luminosity  function of hot DA white dwarfs  is found. With all
these inputs,  we ran our Monte  Carlo code to obtain  a population of
hot DA  white dwarf  models with metal-contaminated  atmospheres.  The
results of  this set  of calculations  are shown in  the top  panel of
Fig.~\ref{f:lf_models}   by  a   dashed  line   connecting  the   open
squares. This synthetic luminosity  function includes both the effects
of NLTE corrections and those of  the metallic contamination of hot DA
atmospheres.  The solid line connecting the filled squares corresponds
to the observed hot DA white  dwarf luminosity function from the SDSS.
As  can be  seen, these  two additional  effects have  a very  limited
impact  on  the  simulated   white  dwarf  luminosity  functions,  and
consequently they  cannot be  at the origin  of the  plateau.  Another
possibility is that for these luminosities the so-called ``Balmer-line
problem'' -- see \cite{Gianinas2010} and references therein -- affects
the  determination of  the  bolometric magnitudes  of  very hot  white
dwarfs in such  a way that the shape of  the luminosity function could
be substantially  modified. To study  this possibility we  adopted the
following  procedure. First,  we  assumed that  a considerable  (20\%)
fraction of  very hot  DA white dwarfs  (namely, those  with effective
temperatures larger  than 40,000~K)  present the  Balmer-line problem.
This fraction is  twice the observed one, but in  this way we maximize
the possible effects of the Balmer-line problem.  Then, to these white
dwarfs we  applied the corrections of  Fig.~12 of \cite{Gianinas2010},
and we recomputed the luminosity  function of DA white dwarfs, keeping
the rest of the inputs the same. The result of this procedure is shown
in the  second panel of  Fig.~\ref{f:lf_models}.  As can be  seen, our
simulations are again  unable to correctly reproduce  the existence of
the plateau in the luminosity function.  Therefore, we now explore our
next possible solution.

\begin{figure}[t]
\vspace{9.0cm}
\includegraphics{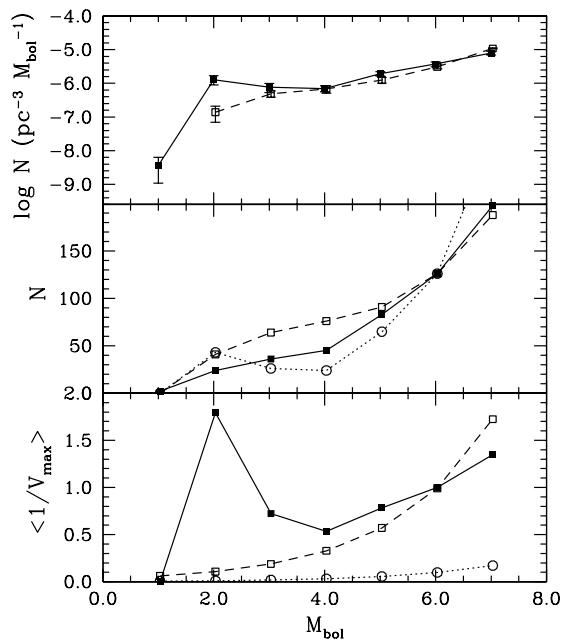}
\caption{Top  panel: the  luminosity function  of DA  white dwarfs  of
  \cite{Krz2009} -- solid line and  filled squares -- and our fiducial
  white   dwarf  luminosity   function   --  dashed   line  and   open
  squares. Middle panel: total number of objects of \citet{Krz2009} --
  solid line and  filled symbols -- and that obtained  in our fiducial
  model --  dashed line and open  squares. In this panel  we also plot
  the antilogarithm of the observed white dwarf luminosity function --
  dotted line  and hollow circles. Bottom  panel: average contribution
  per object,  as obtained from  the data of \citet{Krz2009}  and from
  our  population  synthesis  simulations  --  dashed  line  and  open
  squares.  The  solid line  connecting the  filled squares  shows the
  average  contribution of  DA  white dwarfs,  while  the dotted  line
  connecting the hollow circles shows that of non-DA white dwarfs.}
\label{f:number}
\end{figure}

A second  possibility worth  exploring is  that the  cooling sequences
used to compute the theoretical  luminosity function of the population
of  DA white  dwarfs  are not  reliable for  this  range of  effective
temperatures.  To explore the effects of  a different choice of the DA
white dwarf cooling  tracks, we adopted a completely  different set of
cooling sequences  which incorporate  different prescriptions  for the
model atmospheres from \cite{Tremblay_11}, and have been computed with
very different  stellar evolutionary codes that  incorporate different
physical inputs.   Additionally, we recall  that both sets  of cooling
sequences  --  that  is,  those of  \cite{Fontaine_01}  and  those  of
\cite{Renedo_10}  --   have  been   evolved  from   different  initial
models. The results  of this new set of calculations  are displayed in
the bottom  two panels  of Fig.~\ref{f:lf_models}. As  can be  seen in
this figure,  the impact on  the shape  of the white  dwarf luminosity
function of DA  white dwarfs of the adopted  evolutionary sequences is
minimal.   In  particular,  although using  the  evolutionary  cooling
tracks of \cite{Fontaine_01}  the bins of higher  luminosities -- say,
those with  $M_{\rm bol}\la  1.0$ -- are  more populated,  the overall
shape  of  the  luminosity  function  remains  almost  unchanged,  and
moreover the apparent  plateau of the white  dwarf luminosity function
is not  reproduced, while both  sets of evolutionary sequences  are in
close agreement for  luminosities smaller than this  value.  Hence, we
can  safely discard  the hypothesis  that the  plateau is  an artifact
produced by our  choice of model atmospheres or a  poor description of
the initial phases of white dwarf cooling.

\begin{figure}[t]
\vspace{8.0cm}
\includegraphics{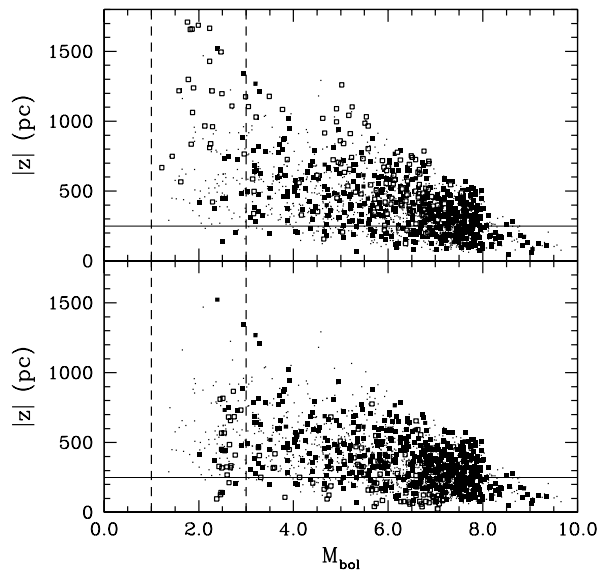}
\caption{Distribution  of distances  above  the Galactic  plane of  DA
  white dwarfs  in the sample  of \cite{Krz2009} and in  our simulated
  sample as a function of  the magnitude. The small symbols correspond
  to the results of our Monte  Carlo simulations, while the large ones
  represent the  observational data. Open symbols  correspond to white
  dwarfs with  derived masses smaller than  $0.49\, M_{\sun}$, whereas
  filled symbols  represent white  dwarfs with measured  masses larger
  than  this  value. The  top  panel  presents  the original  data  of
  \cite{Krz2009},  while  the  bottom panel  shows  revised  distances
  obtained assuming  that the mass  of each  of these white  dwarfs is
  $0.49\, M_{\sun}$.}
\label{f:z}
\end{figure}

Now  we  turn  our  attention  to  the  third  possibility  previously
envisaged -- that  there is an observational bias in  the way in which
the DA white dwarf luminosity  function is computed. In particular, it
may  be possible  that the  observed plateau  is a  consequence of  an
unexpected larger  number of  objects in the  corresponding luminosity
bins.  However, it should be kept in mind that the $1/V_{\max}$ method
is  specifically  designed  to  compensate  for the  bias  due  to  an
observational  magnitude   limit  cut   and  consequently,   that  the
contribution  of  each  individual  star to  the  luminosity  function
depends on its magnitude.  Specifically, the contribution of each star
to  the luminosity  function is  obtained from  the maximum  volume at
which it could be found:
\begin{equation}
V_{\max}=\frac{4}{3}\pi\beta\left(r_{\max}^3-r_{\min}^3\right)
{\rm e}^{-|z|/{\rm H}}
\end{equation}
where $r_{\max}$ and $r_{\min}$ are  the maximum and minimum distances
for which an object with apparent magnitude $m$ can be detected within
the magnitude limits of the survey,  $\beta$ is the sky solid angle of
the sample of stars, $z$ is  the vertical Galactic coordinate, and $H$
is  the Galactic  scale  height,  for which  we  adopt  250~pc, as  in
\cite{Krz2009}.   Therefore, for  a given  absolute magnitude  $M$ the
maximum and minimum distances are a function of the apparent magnitude
cut, so the maximum volume is given by:
\begin{equation}
V_{\max}=\frac{4}{3}\pi\beta10^{-\frac{3}{5}M}
\left(10^{\frac{3}{5}(m_b+5)}-10^{\frac{3}{5}(m_u+5)}\right)
{\rm e}^{-|z|/H}
\end{equation} 
where $m_{\rm u}$ and $m_{\rm b}$  are represent the range of selected
magnitudes.  Note that for a given  magnitude cut, the term inside the
parenthesis is constant. Hence, the  contribution of a given object to
the luminosity function,  which is the inverse of  the maximum volume,
is then:
\begin{equation}
1/V_{\max}\propto 10^{\frac{3}{5}M}{\rm e}^{|z|/H}
\end{equation}
This last equation  implies that the contribution of an  object to the
luminosity function  only depends  on its  absolute magnitude  and its
Galactic height,  $z$. Since  there is no  physical reason  to suspect
that  the  Galactic  scale  height   depends  on  the  luminosity,  or
conversely  that  white  dwarfs (and  their  corresponding  progenitor
stars) born  at different  times come  from Galactic  populations with
different  scale heights,  it ends  up that  the contribution  of each
individual  object to  the  luminosity function  only  depends on  its
absolute magnitude.  Moreover, if it  were the case that  the Galactic
scale height depends  on the luminosity, a similar  feature would also
be present in the luminosity function of non-DA white dwarfs, which is
not the case.

\begin{figure}[t]
\vspace{12.0cm}
\includegraphics{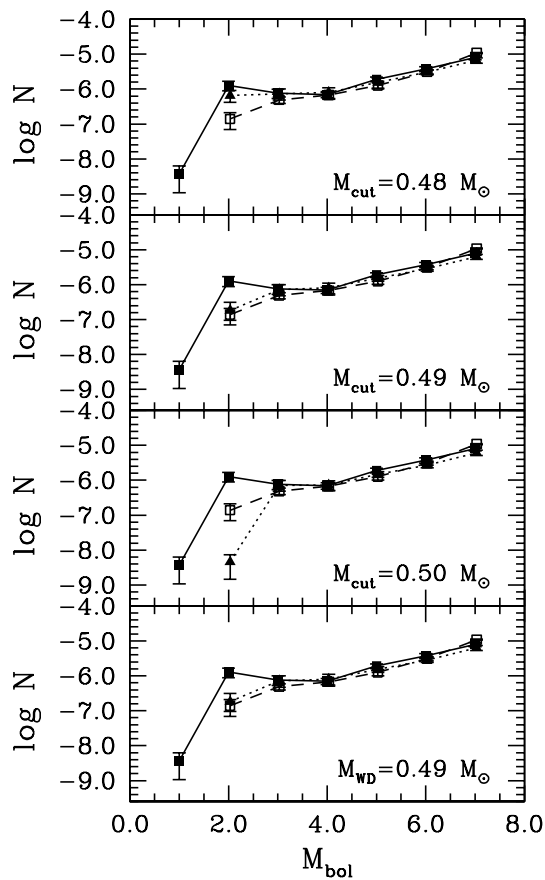}
\caption{Luminosity  functions of  DA  white dwarfs  for several  mass
  cuts.  See text  for details.   The solid  line with  filled squares
  represents the  original \citet{Krz2009}  DA white  dwarf luminosity
  function,  the  dashed  line  with open  squares  is  our  simulated
  luminosity function,  and the dotted  line with filled  triangles is
  the \citet{Krz2009} DA luminosity function with our mass adjustments
  as described in the text.}
\label{f:cuts}
\end{figure}

To  clarify this  issue, Fig.~\ref{f:number}  displays the  luminosity
function of DA white dwarfs in  the SDSS and several other interesting
quantities. In the top panel of this figure, we show both the observed
luminosity   function   (solid   line   with   solid   symbols)   from
\citet{Krz2009} and our fiducial model (dashed line and open symbols).
The middle  panel of Fig.~\ref{f:number}  depicts the total  number of
white dwarfs for each luminosity  bin of the \cite{Krz2009} luminosity
function (solid line  and filled symbols). Note that  the total number
of objects is monotonically decreasing with Luminosity.  Also shown in
this panel  are the total  number of DA  white dwarfs obtained  in our
fiducial  simulation (dashed  line connecting  open squares),  and the
antilogarithm  of the  white  dwarf luminosity  function (dotted  line
connecting  hollow circles)  derived from  the  SDSS --  that is,  the
antilogarithm of the luminosity function shown in the top panel, which
allows  to  gain more  insight  on  the  statistical behavior  of  the
luminosity  function. All  three curves  have been  normalized to  the
luminosity   bin  with   $\log(L/_{\sun})=-0.5$   (that  is,   $M_{\rm
bol}=3.45$).   This  panel  clearly   demonstrates  that  because  our
theoretical  simulations yield  more  white dwarfs  in the  luminosity
range of  the plateau  than does the  observed white  dwarf luminosity
function, their  contribution to  the number density  of stars  in the
region  of the  plateau must  therefore be  smaller.  Moreover,  it is
interesting to note that although the number of objects in the SDSS is
monotonically increasing with  fainter luminosities, the antilogarithm
of the white  dwarf luminosity function is not,  thus pointing towards
an  artifact  existing in  the  observational  white dwarf  luminosity
function.   Finally, in  the bottom  panel of  Fig.~\ref{f:number}, we
show the average $1/V_{\max}$ contribution  for each luminosity bin in
the SDSS sample of DA white dwarfs (solid line and filled squares), in
our fiducial  simulation (dashed  line and open  squares), and  in the
SDSS sample of  non-DA white dwarfs (dotted line  and hollow circles).
These curves  show that  for the  population of  DA white  dwarfs, the
average  value  of  $1/V_{\max}$  presents  a peak  at  the  range  of
appropriate  luminosities  which  is   $\sim  20$  times  larger  than
expected, pointing towards  an artifact produced by one  or several of
the objects  of the  observational sample.   Moreover, given  that our
simulations predict a  larger number of objects in  the relevant range
of effective temperatures, we conclude  that the origin of the plateau
does not reside in an excess  of objects in these luminosity bins, but
rather  in  that their  contribution  to  the luminosity  function  is
somehow larger than it should be.

To investigate  this point in  detail, the top half  of Fig.~\ref{f:z}
shows the  distribution of  distances above the  Galactic plane  of DA
white dwarfs  in the sample of  \cite{Krz2009} -- large symbols  -- as
well as  that obtained in a  typical Monte Carlo realization  -- small
symbols.  Our  simulation matches remarkably well  the distribution of
vertical distances  of the white dwarfs  in the SDSS sample  except in
the region  of the plateau.   which in this  plot is delimited  by the
vertical dashed lines. In particular, there is a group of white dwarfs
in this region  of the SDSS luminosity function  with noticeable large
distances above the Galactic plane that our population synthesis model
does   not  reproduce.   Upon  inspecting   the  original   sample  of
\cite{Krz2009}, we found  that most of these white  dwarfs have masses
well  below $0.49\,  M_{\sun}$ --  the  theoretical lower  limit of  a
carbon-oxygen  white  dwarf  --  and   thus  cannot  be  single  white
dwarfs.  To highlight  the effects  of the  mass determination  in the
luminosity function,  we highlight  these low-mass white  dwarfs using
open squares  whereas those white  dwarfs with masses  compatible with
the  theoretical  lower  limit  for  carbon-oxygen  white  dwarfs  are
displayed using  solid squares.  Clearly,  the vast majority  of white
dwarfs in the region where the  DA white dwarf luminosity function has
a plateau have  determined masses smaller than  $0.49\, M_{\sun}$.  To
explore the effect  of these low mass determinations,  the lower panel
of  Fig.~\ref{f:z} shows  revised distances  above the  Galactic plane
calculated  assuming that  the masses  of  all the  less than  $0.49\,
M_{\sun}$ white dwarfs are instead the minimum mass of a carbon-oxygen
core white dwarf, $0.49\, M_{\sun}$.  Now, the discrepancy between the
observed and simulated samples  largely disappears, and although there
are white  dwarfs with  large distances above  the Galactic  plane the
relative number of these white dwarfs agrees with our simulations.

Fig.~\ref{f:cuts} explores  the effects of the  DA mass determinations
in the  SDSS DA white  dwarf luminosity function.  Since  white dwarfs
with  masses   smaller  than   $0.49\,  M_{\sun}$  cannot   be  single
carbon-oxygen white dwarfs,  either these white dwarfs  are members of
unresolved  binaries  and  thus,  they should  be  excluded  from  the
observational sample, or the mass  determination of these white dwarfs
is  an artifact  of the  reduction procedure,  and therefore  its mass
should  be  at  least the  minimum  mass  for  a  white dwarf  with  a
carbon-oxygen  core.   To  explore  the first  possibility,  we  apply
different mass cuts to the as published SDSS sample and we compute the
DA white  dwarf luminosity  function excluding  all stars  with masses
smaller than the mass cut.  Fig.~\ref{f:cuts} shows the results.  With
a  mass cut  $M_{\rm cut}=0.48\,  M_{\sun}$, the  resulting luminosity
function -- shown  as a dotted line connecting triangles  -- is barely
affected and  is very similar  to the original luminosity  function of
\cite{Krz2009}  -- solid  line connecting  solid squares  -- and  very
different from  our simulated  result --  dashed line  connecting open
squares. However, with  a $M_{\rm cut}=0.49\, M_{\sun}$  mass cut, the
situation changes completely and  the updated observational luminosity
function perfectly agrees  with the simulated one. If the  mass cut is
increased  to $M_{\rm  cut}=0.50\,  M_{\sun}$,  the agreement  between
luminosity functions is  poor, as expected, and  a significant paucity
of hot DA white dwarfs becomes rather evident.  Finally, in the bottom
panel of Fig.~\ref{f:cuts}, we simply assume that DA white dwarfs with
masses  smaller than  $M_{\rm  WD}<0.49\,  M_{\sun}$ have  erroneously
determined  masses and  that  their real  masses  should be  precisely
$0.49\, M_{\sun}$.   Here again,  we see remarkable  agreement between
the simulated and the observational luminosity functions, very similar
to that obtained applying a  mass cut $M_{\rm cut}=0.49\, M_{\sun}$ to
the SDSS sample.  All in all, we conclude that the luminosity function
of DA white dwarfs of \cite{Krz2009} must be revised, since it is very
sensitive to the  mass determinations of DA white  dwarfs.  We provide
the values of the corrected luminosity function in table~\ref{t:wdlf},
where  we list  the bolometric  magnitude, the  space density  and the
number of  stars in each  bin, assuming that  the minimum mass  of the
white dwarfs in the sample of \cite{Krz2009} is $0.49\, M_{\sun}$.

\section{Conclusions}

\begin{table}
\centering
\begin{tabular}{ccc}
\hline
\hline
$M_{\rm bol}$ & $\log N$~(pc$^{-3}$~$M_{\rm bol}^{-1}$) & Number of stars \\
\hline 
\noalign{\smallskip}
2    & $-6.937^{+0.230}_{-0.284}$ &  12 \\ 
\noalign{\smallskip}
3    & $-6.342^{+0.138}_{-0.179}$ &  45 \\ 
\noalign{\smallskip}
4    & $-6.326^{+0.139}_{-0.173}$ &  47 \\ 
\noalign{\smallskip}
5    & $-6.018^{+0.076}_{-0.089}$ &  75 \\ 
\noalign{\smallskip}
6    & $-5.739^{+0.063}_{-0.072}$ & 115 \\ 
\noalign{\smallskip}
7    & $-5.375^{+0.067}_{-0.079}$ & 209 \\ 
\noalign{\smallskip}
\hline
\hline
\end{tabular}
\caption{Observed  DA white  dwarf  luminosity  function dwarfs  after
  applying  the  correction  for   those  objects  with  masses  below
  $0.49\,M_{\sun}$.}
\label{t:wdlf}
\end{table}

We presented  a set of simulations  of the luminosity function  of hot
white  dwarfs.   Our  results  are in  excellent  agreement  with  the
observations for non-DA stars, while for DA white dwarfs we found that
a plateau  present in the  luminosity function at  $M_{\rm bol}\approx
2.0$ --  or, equivalently, $\log  (L/L_{\sun})\approx 1$ --  cannot be
reproduced by  our simulations. We  first discarded that  this plateau
could be due to  a global Galactic input, like a  recent burst of star
formation, because if  this were the case, the  luminosity function of
hot  non-DA white  dwarfs  should  show a  similar  plateau.  We  then
investigated other  possible origins of this  plateau.  In particular,
we studied the  possible role of NLTE effects  and metal contamination
in  the atmospheres  of hydrogen-rich  white dwarfs,  the role  of the
so-called  Balmer-line problem  for  very hot  white  dwarfs, and  the
possibility  that  the  available   cooling  sequences  were  severely
affected by  the initial conditions  adopted to compute them,  and the
statistical uncertainties in the white  dwarf sample. We found that at
the  relevant  luminosities  of  the  plateau,  the  effects  of  NLTE
corrections are  limited and  that metal  contamination plays  a minor
role in shaping  the luminosity function.  This is also  the case when
the Balmer-line problem is taken  into account.  We also discarded the
explanation that the available cooling tracks are severely affected by
the choice of initial conditions.  Thus,  we were left with a possible
statistical fluctuation,  which we  investigated in detail.   We found
that even though  our theoretical simulations yield  more white dwarfs
in  the   luminosity  range   of  the  plateau   than  exist   in  the
\citet{Krz2009} sample,  their contribution  to the number  density of
stars in  this region is  smaller than that  of the SDSS  sample, thus
pointing towards an artifact in  the derivation of the SDSS luminosity
function.

We therefore analyzed the distribution  of masses and distances of the
SDSS sample and we found that  within the range of luminosities of the
plateau, the SDSS white dwarf luminosity function is almost completely
dominated  by the  contributions of  a  handful of  white dwarfs  with
masses smaller than $0.49\, M_{\sun}$,  the lower theoretical mass for
a carbon-oxygen white dwarf. Thus, either these objects are members of
unresolved binaries,  or they have erroneously  determined masses.  We
found that once these objects are excluded from the calculation of the
DA white dwarf  luminosity function, the agreement  with our simulated
luminosity  function  is  excellent.  Additionally,  we  computed  the
distances setting the mass of each  of these $< 0.49\, M_{\sun}$ white
dwarfs at  $0.49\, M_{\sun}$,  again finding excellent  agreement with
our simulated  luminosity functions and  that obtained with  a $0.49\,
M_{\sun}$ mass cut applied to  the SDSS sample.  We therefore conclude
that the luminosity function of  hot DA white dwarfs of \cite{Krz2009}
is contaminated by white dwarfs with incorrectly determined low masses
(below  $0.49\,  M_{\sun}$)  and  we provide  a  revised  white  dwarf
luminosity function in Table~\ref{t:wdlf}.   The analysis presented in
this paper also  confirms the validity of the  white dwarf theoretical
cooling sequences at high effective temperatures via the good match of
our model  luminosity functions to  that of the SDSS  observations. We
also  emphasize that  the way  in which  the observational  luminosity
function of  hot white  dwarfs is  obtained is  very sensitive  to the
particular implementation of  the method used to derive  the masses of
the sample.

\begin{acknowledgements}
This research was  supported by AGAUR, by  MCINN grant AYA2011--23102,
by the European Union FEDER funds,  and by the ESF EUROGENESIS project
(grant EUI2009-04167).
\end{acknowledgements}

\bibliographystyle{aa}
\bibliography{hots}

\end{document}